\documentclass[12pt]{article}
\usepackage{amsmath}
\usepackage{amssymb}
\usepackage{graphicx}
\usepackage{hyperref}
\usepackage{enumitem}

\begin{document}

\numberwithin{equation}{section}

\begin{titlepage}
\hbox to \hsize{\hspace*{0 cm}\hbox{\tt }\hss
   \hbox{\small{\tt }}}

\vspace{1 cm}

\centerline{\bf \Large Operator mixing in deformed D1D5 CFT} 
\bigskip
\centerline{\bf \Large and the OPE on the cover}

\vspace{1 cm}

\vspace{1 cm}
\centerline{\large Benjamin A. Burrington $^{\star}$\footnote{benjamin.a.burrington@hofstra.edu} ,
Ian T. Jardine$^\dagger$\footnote{jardinei@physics.utoronto.ca}\,, and
Amanda W. Peet$^{\dagger\S}$\footnote{awpeet@physics.utoronto.ca}}

\vspace{0.5cm}

\centerline{\it ${}^\star\!\!$ Department of Physics and Astronomy, Hofstra University, Hempstead, NY 11549, USA}
\centerline{\it ${}^\dagger$Department of Physics, University of Toronto, Toronto, ON M5S 1A7, Canada}
\centerline{\it ${}^\S$Department of Mathematics, University of Toronto, Toronto, ON M5S 2E4, Canada}

\vspace{0.3 cm}

\begin{abstract}

We consider the D1D5 CFT near the orbifold point and develop methods for computing the mixing of untwisted operators to first order by using the OPE on the covering surface. We argue that the OPE on the cover encodes both the structure constants for the orbifold CFT and the explicit form of the mixing operators. We show this explicitly for some example operators.  We start by considering a family of operators dual to supergravity modes, and show that the OPE implies that there is no shift in the anomalous dimension to first order, as expected.  We specialize to the operator dual to the dilaton, and show that the leading order singularity in the OPE reproduces the correct structure constant.  Finally, we consider an unprotected operator of conformal dimension (2,2), and show that the leading order singularity and one of the subleading singularies both reproduce the correct structure constant. We check that the operator produced at subleading order using the OPE method is correct by calculating a number of three point functions using a Mathematica package we developed. Further development of this OPE technique should lead to more efficient calculations for the D1D5 CFT perturbed away from the orbifold point.

\end{abstract}

\end{titlepage}

\tableofcontents

\section{Introduction}
One of the most important questions in quantum gravity is the black hole information problem. Although there have been many proposals for correcting Hawking's picture semiclassically, Mathur made a broad-based argument that more major quantum gravity corrections will be needed to resolve the information problem \cite{Mathur:2009hf}. Firewall arguments of Almheiri et al considerably sharpened community thinking about the regime of validity of effective field theories involving Einstein gravity \cite{Almheiri:2012rt}. Proposals for rescuing ourselves from information loss have included ER=EPR wormholes \cite{Maldacena:2013xja}, state-dependent operators \cite{Papadodimas:2013jku}, and non-locality both in a semiclassical setting \cite{Giddings:2012gc} and arising from string effects near horizons \cite{Silverstein:2014yza,Dodelson:2015toa}.

Another interesting approach has been the fuzzball program (see \cite{Mathur:2005zp} for an introduction). Its general framing is more top-down than bottom-up, and it focuses on black hole microstates, whose geometries are string theoretic constructions without horizons or singularities. The general goal is for the sum over microstates to capture the full physics of the black hole. For any given microstate, the infinite throat geometry a classical relativist would expect is replaced by a long but finite throat, ending in an intricate, distinct string theoretic arrangement of fluxes, compactified dimensions, and nonperturbative ingredients. Order one corrections set in just before the would-be horizon, with no degrees of freedom lost. Classes of such microstates have been well studied, with  black hole entropy and radiation emission were reproduced beautifully \cite{Strominger:1996sh}\cite{Avery:2009tu}. Fuzzball microstates can also provide order one changes to the entanglement structure needed to resolve the black hole information problem \cite{Mathur:2012jk}.

The holographic AdS/CFT correspondence \cite{Maldacena:1997re} has been a  fruitful laboratory for exploring many ideas involving quantum gravity, including the black hole information problem. In this framework, dynamics of quantum gravity in AdS are described in terms of a conformal field theory (CFT) without gravity living in one fewer dimensions. This duality is strong-weak, meaning that strongly coupled AdS quantum gravity can be studied via weakly coupled CFT and vice versa. Certain supersymmetric quantities are protected by non-renormalization theorems, allowing comparisons of two weak coupling calculations from either side \cite{Lee:1998bxa}.
Recent advances in CFT technology have enabled asking questions about the space of CFTs that are holographic, and about what features of holographic CFTs might be universal in $1/c$ and $h/c$ expansions, such as in
\cite{Hartman:2014oaa,Keller:2014xba,Perlmutter:2015iya,Belin:2016yll,Anous:2016kss,Fitzpatrick:2016ive,Chen:2016cms}.

A well-studied prototype for black holes in string theory is the D1D5 system, constructed using D1 and D5 branes wrapped on $T^4\times S^1$. In the low energy regime, the dual geometry has a supergravity description in terms of a five-dimensional supersymmetric black hole with a degenerate horizon. At another point in its moduli space, the system's dual CFT has a description as a free $(T^4)^N/S_N$ orbifold CFT. Periodicity up to the action of the orbifold allows new boundary conditions and introduces twisted sectors to the theory, and associated to these new states are new operators, the twist operators.
However, astrophysical black holes are obviously not supersymmetric, and so there have been a number of works dedicated to constructing more complicated and physical microstate geometries \cite{Mathur:2013nja,Bena:2013dka,Bena:2015bea,Bena:2016agb}. A number of these constructions are three charge systems with CFT duals which are states in the D1D5 CFT. So understanding the D1D5 CFT is important for understanding even more complicated constructions in the fuzzball program.

The D1D5 system has a moduli space; the supergravity description and the free orbifold description lie at two different points in it.  A natural question that arises is how to deform the free orbifold CFT towards the supergravity limit.  This is accomplished by one of the marginal operators in the theory, the deformation operator \cite{David:1999ec}. The deformation operator has been studied in a wide range of contexts \cite{Gava:2002xb,Avery:2010er,Avery:2010hs,Burrington:2012yq,Carson:2014yxa,Carson:2014xwa,Carson:2014ena,Burrington:2014yia,Gaberdiel:2015uca,Carson:2015ohj,Carson:2016cjj,Carson:2016uwf}.  This operator belongs to the twisted sector of the theory, and is the primary motivation for our current investigation.

Working with the D1D5 CFT requires unique techniques to deal with operators in the twisted sector. One of the most important of these is the Lunin-Mathur technique \cite{Lunin:2000yv,Lunin:2001pw}, where one lifts operators and correlators to a covering space where the twisted boundary conditions are encoded by ramified points in the map to the cover. Numerous applications have including matching correlators to bulk calculations (e.g. emission rates \cite{Avery:2009tu}), information loss \cite{Galliani:2016cai} , entanglement entropies and other related quantities \cite{Headrick:2010zt,Lin:2016fqk}, and also thermalization \cite{Avery:2010er,Carson:2014ena}. We can also apply other helpful techniques, such as bosonization, to simplify D1D5 CFT calculations  \cite{Burrington:2015mfa}.  Despite this large body of work, the full description of the D1D5 CFT, especially away from the free orbifold point, is still not fully understood.

Here we continue to develop the covering space techniques of \cite{Burrington:2012yn,Burrington:2012yq} to learn about conformally perturbing away from the orbifold point.  In our earlier work \cite{Burrington:2012yq} we considered certain four point functions which contain two copies of our candidate operator and two copies of the deformation operator.  Taking a coincidence limit of such a four point function allowed us to read off structure constants and show that at first order in perturbation theory a specific non-protected operator mixed with an unknown operator of the same conformal dimension, and in this case both operators acquire an anomalous dimension.  One would have to find the exact form of the operator that mixes with the candidate operator, and iterate this procedure to find all operators that mix in this block.  At each stage a four point function could be computed to find if there was a new operator that mixes, and if so, many three point functions would need to be computed to find out the proper combination that mixes.  This is a fairly straightforward but laborious technique.

In this paper, we take another approach.  The covering space techniques focus on the computation of $n$-point functions, particularly three and four point functions.  The three point functions of quasi-primaries give the structure constants.  The structure constants are the main building blocks for any CFT, and using bootstrap techniques allows the building of general correlation functions.  There is of course another mainstay of CFT that contains the information about the structure constants: the operator product expansion (OPE).  The symmetric orbifold CFT on the covering surface is particularly simple: it is a free theory.

Thus, from previous work, we may say the following:
\setlist{nolistsep} 
\begin{itemize}[noitemsep] 
\item If we know what the lift of an operator is to the cover, then we know everything about the operator.  We have enough to compute any correlator.  Further, if we can find it for one map, then we can reconstruct it for any map by reexpressing the operator on the cover as appropriate modes of fields acting on twists that would lead to that particular operator for that particular map.
\item The operator product expansion contains all information about three point functions because the OPE is written in terms of the structure constants.
\item The OPE of free theories is simple.  The free orbifold theory is a little more complicated, but on the covering surface the difficulties of boundary conditions are removed.  The twisted boundary conditions are accounted for in the map from the base space to the covering surface.  Thus, the OPE on the cover is simple.
\end{itemize}
Given these statements, there is a natural question:  Does the knowledge of the OPE on the covering surface descend in some way to knowledge of the OPE on the base?

It is the purpose of this paper to offer evidence that the answer to this question is ``yes,'' and outline how to do this in the special case where one of the operators is in the untwisted sector. We do this by considering specific operators as examples.  We lay out the challenges of general twist-twist OPEs in the Discussion.

The eventual goal of these techniques would be to work out all the mixing operators for a candidate operator and then use conformal perturbation theory to compute the anomalous dimensions of the operators after the deformation. This would illuminate how the free CFT point is connected to the supergravity point in moduli space.  Using the the OPE on the cover and simply reading off the operator at the next stage of mixing is much more efficient than computing the plethora of three point functions at each stage.

The layout of the paper is as follows. Section \ref{D1D5CFT} contains details about the D1D5 CFT and the techniques used to compute the structure constants in the paper. Section \ref{SUGRAoperator} will show using the OPE and the lift to the cover that the exactly marginal supergravity operators do not have an anomalous dimension, as expected. We will also show how one can reproduce the leading singularity of the four point function for the dilaton operator and deformation operator with the OPE. Section \ref{stringoperator} will use the OPE method to find the operator that mixes with a (2,2) untwisted operator and contributes to its anomalous dimension. We will also show how this result can be verified by computing three point functions and confirm these results by reproducing the coefficient for the mixing found in the four point function. We finish with the Discussion, including some directions for future work, in Section \ref{discussion}.

\section{D1D5 CFT and the Lunin-Mathur method}\label{D1D5CFT}
\subsection{D1D5 CFT review}
The D1D5 CFT is a well known CFT describing the D1D5 system, see \cite{Seiberg:1999xz} \cite{Vafa:1995bm} \cite{Larsen:1999uk} \cite{deBoer:1998ip} \cite{Dijkgraaf:1998gf} \cite{Arutyunov:1997gt} \cite{Arutyunov:1997gi} \cite{Jevicki:1998bm} for some of the earlier papers on this system. It can be described by a $N=(4,4)$ 1+1d free SCFT with the target space $(T^4)^{N_1N_5}/S_{N_1N_5}$ at a point in its moduli space. Here $N_1$ is the number of D1 branes and $N_5$ is the number of D5 branes. This SCFT has R symmetry $SU(2)_L\times SU(2)_R$ and another custodial symmetry inherited by the symmetry of the $T^4$, $SO(4)\sim SU(2)_1\times SU(2)_2$. Each copy has scalar fields which are given by $X_{i}$ but we rewrite them as,
\begin{equation}
X^{\dot{A}A}=\frac{1}{\sqrt{2}}(\sigma^i)^{\dot{A}A}X_i,
\end{equation}
where $A,\dot{A}$ are the indicies of the $SU(2)_1\times SU(2)_2$. We write the fermions as $\psi^{\alpha\dot{A}}$ and $\tilde{\psi}^{\dot{\alpha}\dot{A}}$ where $\alpha,\dot{\alpha}$ are the indicates of the $SU(2)_L\times SU(2)_R$, respectively. These complex fermions satisfy a hermiticity constraint,
\begin{equation}
(\psi^{\alpha\dot{A}})^{\dagger}=-\epsilon_{\alpha\beta}\epsilon_{\dot{A}\dot{B}}\psi^{\beta\dot{B}}
\end{equation}
and similarly for the other fields. Our conventions for the epsilon tensors will be that $\epsilon_{12}=1=\epsilon^{21}$, with the Pauli matrices $(\sigma^{2})^{\dot{2}1}=i$. In addition to these, we also have definitions for the supercharge, current, and energy momentum tensor. These are given by, respectively,
\begin{align}
&G^{\alpha A}=\epsilon_{\dot{A}\dot{B}}\psi^{\alpha\dot{A}}\partial X^{\dot{B}A},\\
&J^{a}=\frac{1}{4}\epsilon_{\dot{A}\dot{B}}\epsilon_{\alpha\beta}(\sigma^{*a})^{\beta}_{\gamma}\psi^{\alpha\dot{A}}\psi^{\gamma\dot{B}},\\
&T=\frac{1}{2}\epsilon_{AB}\epsilon_{\dot{A}\dot{B}}\partial X^{\dot{A}A}\partial X^{\dot{B}B}+\frac{1}{2}\epsilon_{\dot{A}\dot{B}}\epsilon_{\alpha\beta}\psi^{\alpha\dot{A}}\partial\psi^{\beta\dot{B}},
\end{align}
with similar expressions for the anti-holomorphic side. These satisfy various commutation relations which can be found in \cite{Avery:2010qw}.

In addition to the above fields we also have a twisted sector from the action of the orbifold. Operators in the twisted sector involve a bare twist $\sigma_{(12...n)}=\sigma_n$, which enforce a twisted boundary condition for fields at the insertion point which permutes the copies of the CFT. These twist fields allow construction of fractionally moded operators,
\begin{equation}
\mathcal{O}_{-m/n}=\oint\frac{dz}{2\pi i}\sum\limits_{k=1}^{n}\mathcal{O}_{-m,(k)}e^{-2\pi i m (k-1)/n}z^{\Delta-m/n-1}.
\end{equation}
These are only well defined in the presence of the twist $n$ operator, so that the integral is periodic around the insertion. Fractional $J^a$, together with spin fields for even $n$, can be used to create twisted chiral primaries. Well defined operators must be $S_N$ invariant and so these twisted operators must be summed over the conjugacy class of the representative permutation given in the bare twist. In general, this will only give an overall combinatoric factor. However, the $S_N$ invariance is important in restricting the form of correlators with fractional weights. We will provide some details in the Discussion after the key calculations are presented.

The D1D5 system is described by a supergravity solution and a free CFT, but these are at different points at the moduli space. The base system has 25 moduli, 5 of which are fixed by the attractor mechanism. This leaves 20 moduli, corresponding to 20 marginal operators. There are four of these in the twisted sector, corresponding to various blow up modes in the bulk. We wish to study the marginal deformation operator which is a singlet of the various $SU(2)$s; this particular blow up mode corresponds in the bulk to $a_1C_{0}+a_2C_{ijkl}$. This will involve computing correlators with operators in the twisted sector. More details of the moduli space can be found in a comprehensive review \cite{David:2002wn}.

The deformation operator in the CFT takes the form of
\begin{equation}\label{defoperator}
\mathcal{O}_{D}=\epsilon_{AB}\epsilon_{\alpha\beta}\epsilon_{\dot{\alpha}\dot{\beta}}G^{\alpha A}_{-1/2}\tilde{G}^{\dot{\alpha} B}_{-1/2}\sigma^{\beta\dot{\beta}}_2.
\end{equation}
This is a singlet under the SU(2)s and is a twist 2 operator. Conformal perturbation with this twisted operator will require using the Lunin-Mathur technique.

\subsection{Lunin-Mathur and bosonization techniques}

To calculate correlations functions involving correlation functions of operators in the twisted sector involves splitting the correlator into the contribution form the non-twisted part and the bare twist operator. The bare twist operator contribution is calculated using the technique first introduced by Lunin and Mathur\cite{Lunin:2000yv,Lunin:2001pw}. We will briefly review it here.

To compute the contribution of these bare twist operators, it is easier to transform to a covering space where these operators are brought to identities (i.e. the boundary conditions are untwisted). This map is called the lifting map and this process is called lifting. To compute the contribution on the cover, we will have to regulate around infinities and around the twist insertions themselves. This leads to  ``pizza diagrams" which help identify the various regulation patches and their copies on the covering space.

This lifting procedure comes with a complication. The D1D5 CFT has $c=6N_1N_5\neq0$ and so, when we lift, we must keep track of the contribution from the conformal anomaly to correlation functions. This is done by introducing a Liouville field. The action of the Liouville field will give three distinct contributions to the correlation function. Two of these are just regulation dependent and cancel with proper normalization of the correlation functions. The important term comes from boundary of the holes cut around the twist insertions and can be calculated using the Liouville action.

This contribution together with the contribution of the non-twist fields will yield the full correlation function. More details and explicit calculations can be found in \cite{Lunin:2000yv}\cite{Lunin:2001pw}. We can use the results from \cite{Avery:2010qw} Appendix D, which has general expressions for the bare twist contributions using the Lunin-Mathur technique.

Another important aspect to computing correlators is bosonization of the fermions. In general, it is easier to compute correlators of bosons instead of fermions. However, fermions have anti-commuting properties not captured by bosons and they need to be enforced using cocycles. This has been studied extensively for the D1D5 system in \cite{Burrington:2015mfa}; here, we will only mention results that we require.

Calculating correlators with spin and fermion fields can be difficult. One way of simplifying these calculations is by bosonizing these fermionic fields into the form,
\begin{equation}
e^{k\cdot\Phi}C_{k},
\end{equation}
where $C_k$ is the cocycle. It is responsible with restoring the fermionic statistics of the fields that it was derived from. We will use the notation and conventions from \cite{Burrington:2015mfa}. We take $C_k=\exp(i\pi k\cdot (M\alpha_0))$, where $\alpha_0$ are momentum operators and $M$ is a matrix of constants which are constrained by the anticommutation relations and OPEs of the fermions and spin fields. We also have $k\cdot\Phi=k_5\phi_5+k_6\phi_6+k_{\tilde{5}}\tilde{\phi}_5+k_{\tilde{6}}\tilde{\phi}_6$ where the tildes represent the antiholomorphic bosonic fields. The commutation for the momentum and the fields is $[\phi_k,\alpha^j]=i\delta^{j}_k$. The paper cited above then provides lists of the bosonized forms of $\psi, S$, and $J$.

It also provides some constraints that must be satisfied by the matrix $M$ for this to work out. These come from enforcing the antisymmetric conditions on the fermions and the OPE structure of the spin fields. For example,
\begin{align}
\psi^{+\dot{1}}(z)\psi^{+\dot{2}}(w)&=e^{-i\pi M_{6i}\alpha_0^i}(e^{-i\phi_6})(z)e^{i\pi(M_{5i})\alpha_0^i}(e^{i\phi_5})(w),\\
&=e^{\pi[\phi_6,M_{5i}\alpha_0^i]}e^{\pi[M_{6i}\alpha_0^i,\phi_5]}e^{i\pi M_{5i}\alpha_0^i}(e^{i\phi_5})(w)e^{i\pi M_{6i}\alpha_0^i}(e^{-i\phi_6})(z),\\
&=e^{i\pi(M_{56}-M_{65})}\psi^{+\dot{2}}(w)\psi^{+\dot{1}}(z),
\end{align}
since we wish for these to anticommute, we then require that $M_{56}-M_{65}\equiv1$ (mod 2). When all the anticommutation relations and OPE structures for the spin fields are accounted for, the conditions are $A_{ij}\pm2\equiv-A_{ij}$ (mod 4), where $A_{ij}=M_{ij}-M_{ji}$. For this work, we took the solution of these constraints as $(A_{56},A_{5\tilde{5}},A_{5\tilde{6}},A_{6\tilde{5}},A_{6\tilde{6}},A_{\tilde{5}\tilde{6}})=(-1,-1,1,1,1,1)$.

Now that we have the techniques in hand to compute correlators, we must decide how to extract the physics of the perturbation by the deformation operator.

\subsection{Structure constants}

In this paper, we wish to study mixing of operators with the future goal of computing anomalous dimensions. We will use conformal perturbation theory to find these anomalous dimensions.  This was discussed in more details in the series of papers \cite{Burrington:2012yn}\cite{Burrington:2012yq}. Where there is no degeneracy for the fields in terms of conformal weights, the anomalous dimension to first order in the deformation, $\mathcal{O}_D$ for an operator $\mathcal{O}_i$ is
\begin{equation}
\frac{\partial h_i}{\partial\lambda}=-\pi C_{iDi},\quad
\frac{\partial \tilde h_i}{\partial\lambda}=-\pi C_{iDi}.
\end{equation}

Here $C_{iDi}$ is the structure constant corresponding to the three point function of two $\mathcal{O}_i$ and $\mathcal{O}_D$. To lift the restriction of no degeneracy, we must diagonalize $C_{iDk}$ in the entire block of fields with same conformal dimension. These mixing operators are what participate in the computations of the anomalous dimension. Other operators with non-vanishing structure constants will only end up affecting the wavefunction renormalization.

So the challenge for finding anomalous dimensions is to compute the structure constants $C_{iDk}$ and the operators $\mathcal{O}_k$, as was alluded to in the Introduction. To do this through three point functions, we must first consider all operators that would possibly participate in the mixing. This means that they first must respect any global symmetry constraints and have the correct conformal weights. Then each of these operators $\mathcal{O}_k$ would then need to be put into the three point function $\langle\mathcal{O}_i\mathcal{O}_D\mathcal{O}_k\rangle$ and computed. If this vanishes, then the operator does not mix and if it is non-zero then in must be included in the block diagonalization. These mixing operators would themselves mix with other operators, $\mathcal{O}_l$ and these also must be included to complete the diagonlization. This would continue until all operators have been found. This process would be tedious to do by hand. To help with the computations, we developed a Mathematica package to help with lifting, determining candidate mixing operators, and computing correlation functions using conformal Wick contractions. We ran some of the computations using the SciNet cluster located at the University of Toronto \cite{SciNetCitation}.

However, this is a very lengthy process and moreover, there are a lot of operators that would have no mixing. So there is a lot of time spent computing vanishing three point functions. Instead, we can consider using the OPE. Recall the basic definition of the OPE,
\begin{equation}
\mathcal{O}_i(z,\bar{z})\mathcal{O}_D(w,\bar{w})=\sum_k(z-w)^{h_k-h_i-h_D}(\bar{z}-\bar{w})^{\bar{h}_k-\bar{h}_i-\bar{h}_D}C_{iDk}\mathcal{O}_k(w,\bar{w})
\end{equation}
So if we could compute the OPE of the deformation operator and an operator, we could extract both the full mixing operator and structure constant directly without having to compute a large number of three point functions. If we further specialize to $h_i=h_k$, $\bar{h}_i=\bar{h}_k$, and recalling that $h_D=\bar{h}_D=1$, then we would only need the one term in the OPE,
\begin{equation}
|z-w|^{-2}C_{iDk}\mathcal{O}_k(w,\bar{w})
\end{equation}
So instead of computing a multitude of three point functions we can instead just compute the OPE and get the one operator that accounts for the full mixing, $\mathcal{O}_k$ and get the required structure constant as well. One would still need to iterate this procedure to find all mixing operators, but this still cuts down significantly on the computational workload.

With these two approaches explained, we turn to showing how these would work with a simpler example in the next Section before getting into a less trivial calculation in Section \ref{stringoperator}.

\section{Results for supergravity operator}\label{SUGRAoperator}
\subsection{OPE and lack of anomalous dimension}

In order to compute structure constants, we must focus in on specific operators. To start off, we take a quick look at how we can show how the SUGRA modes do not have any mixing that contributes to their anomalous dimensions. So consider the OPE,
\begin{equation}
\mathcal{O}_s(z,\bar{z})\mathcal{O}_D(0,0)
\end{equation}
Where $\mathcal{O}_D$ is the deformation operator and we have the general SUGRA mode,
\begin{equation}
\sum\limits_i\partial X_{\dot{C}C,(i)}(z)\bar{\partial}X_{\dot{D}D,(i)}(\bar{z})
\end{equation}
Note we do not sum over the copies, and instead keep them general. These correspond to SUGRA modes in the D1D5, as shown in equation 6.16 of \cite{David:2002wn}.

 Now we consider lifting this operator. Since we are lifting copies separately, we have to be careful about the patches on the double cover and choices of branches. We are lifting with $z=-t^2$, then we lift with copy (1) having $t_1=\sqrt{-z}$ and copy (2) in the patch defined by $t_2=-\sqrt{-z}$. So we can lift the operator as
\begin{align}
\sum\limits_i\partial X_{\dot{C}C,(i)}(z)\bar{\partial}X_{\dot{D}D,(i)}(\bar{z})&\rightarrow\sum\limits_i\left(\frac{dz}{dt}\right)^{-1}(t_i)\left(\frac{d\bar{z}}{d\bar{t}}\right)^{-1}(\bar{t}_i)\partial X_{\dot{C}C}(t_i)\bar{\partial}X_{\dot{D}D}(\bar{t}_i)\\
&= \sum\limits_i\frac{1}{4t_i\bar{t}_i}\partial X_{\dot{C}C}(t_i)\bar{\partial}X_{\dot{D}D}(\bar{t}_i)
\end{align}
Then we can use the fact that $\mathcal{O}_D\rightarrow-i\epsilon^{AB}\partial X_{\dot{A}A}\bar{\partial}X_{\dot{B}B}S^{\dot{A}\dot{B}}$. Then we can put these together and find
\begin{align}
&\sum\limits_i\frac{1}{4t_i\bar{t}_i}\partial X_{\dot{C}C}(t_i)\bar{\partial}X_{\dot{D}D}(\bar{t}_i)(-i\epsilon^{AB}\partial X_{\dot{A}A}\bar{\partial}X_{\dot{B}B}S^{\dot{A}\dot{B}})\\\nonumber
&=\sum\limits_i(-\epsilon^{AB}\frac{1}{4t_i\bar{t}_i}(\frac{\epsilon_{\dot{C}\dot{A}}\epsilon_{CA}}{t_i^2}+\partial X_{\dot{C}C}(t_i)\partial X_{\dot{A}A})(\frac{\epsilon_{\dot{D}\dot{B}}\epsilon_{DB}}{\bar{t}_i^2}+\bar{\partial}X_{\dot{D}D}(\bar{t}_i)\bar{\partial}X_{\dot{B}B})S^{\dot{A}\dot{B}})\\
&=\sum\limits_i(-\frac{i}{4}\epsilon^{AB}(\frac{\epsilon_{\dot{C}\dot{A}}\epsilon_{CA}}{t_i^3}+\frac{1}{t_i}\partial X_{\dot{C}C}(t_i)\partial X_{\dot{A}A})(\frac{\epsilon_{\dot{D}\dot{B}}\epsilon_{DB}}{\bar{t}_j^3}+\frac{1}{\bar{t}_j}\bar{\partial}X_{\dot{D}D}(\bar{t}_j)\bar{\partial}X_{\dot{B}B})S^{\dot{A}\dot{B}})
\end{align}
We can immediately see that we do not have anything that goes like $z^{-1}\bar{z}^{-1}=t^{-2}\bar{t}^{-2}$. The most singular term, $t^{-3}$, does not have any Taylor expansion to make it into a $t^{-2}$ and the rest is too non singular to contribute. So then we have no mixing that contributes to anomalous dimensions, as we expect.

We can also see that the OPE is non-zero, so there is still mixing. These contributions will not affect the anomalous dimensions, but will affect the wavefunction renormalization to first order in the deformation. We will show in the next Section how this will work.

\subsection{OPE and wavefunction mixing}

For the moment, let us specialize to the dilaton. The four point function of two dilaton and two deformations was found in \cite{Burrington:2012yq}. We can now try to reproduce the leading singularity to provide a non-trivial check on the OPE calculation in the previous subsection. The dilaton is found from the general SUGRA operator when there is a simple contraction of the $SU(2)$ indicies. Then we have
\begin{equation}
\mathcal{O}_{dil}(z,\bar{z})=-\epsilon^{\dot{C}\dot{D}}\epsilon^{CD}\sum\limits_i\partial X_{\dot{C}C,(i)}(z)\bar{\partial}X_{\dot{D}D,(i)}(\bar{z})
\end{equation}
The OPE then becomes
\begin{align}
&\mathcal{O}_{dil}(z,\bar{z})\mathcal{O}_D(0,0)\\\nonumber
&\rightarrow\frac{i}{2}\epsilon^{\dot{C}\dot{D}}\epsilon^{CD}\epsilon^{AB}(\frac{\epsilon_{\dot{C}\dot{A}}\epsilon_{CA}}{t^3}+\frac{1}{t}\partial X_{\dot{C}C}(t)\partial X_{\dot{A}A})(\frac{\epsilon_{\dot{D}\dot{B}}\epsilon_{DB}}{\bar{t}^3}+\frac{1}{\bar{t}}\bar{\partial}X_{\dot{D}D}(\bar{t})\bar{\partial}X_{\dot{B}B})S^{\dot{A}\dot{B}}
\end{align}
Note here that the full sum over copies will only involve two copies for any representative twist we might consider, so we end up with a factor of two for any term where we do not have a difference in sign. In general, one should do the Taylor series first then do the summing over images. This would lead so some terms canceling in the end. For our interests, this subtlety will not affect our results. So for now, the negative signs that come from the differing patches can be ignored, as the holomorphic and antiholomorphic as summed together and so the signs end up canceling and we can write it in terms of a generic $t$. Continuing, we can expand this out and simplify
\begin{align}
&\mathcal{O}_{dil}(z,\bar{z})\mathcal{O}_D(0,0)\\\nonumber
&\rightarrow\frac{i}{|t|^6}\epsilon_{\dot{A}\dot{B}}S^{\dot{A}\dot{B}}+\frac{i}{2\bar{t}^2|t|}\epsilon^{AB}\partial X_{\dot{B}B}(t)\partial X_{\dot{A}A}S^{\dot{A}\dot{B}}+\epsilon^{AB}\frac{i}{2t^2|t|}\bar{\partial}X_{\dot{A}A}(\bar{t})\bar{\partial}X_{\dot{B}B}S^{\dot{A}\dot{B}}\\\nonumber
&+\frac{i}{2|t|}\epsilon^{AB}\epsilon^{\dot{C}\dot{D}}\epsilon^{CD}\partial X_{\dot{C}C}(t)\partial X_{\dot{A}A}\bar{\partial}X_{\dot{D}D}(\bar{t})\bar{\partial}X_{\dot{B}B}S^{\dot{A}\dot{B}}
\end{align}
Now let us consider just the first term,
\begin{equation}
\frac{i}{|t|^6}\epsilon_{\dot{A}\dot{B}}S^{\dot{A}\dot{B}}
\end{equation}
We can write this in terms of a power series in $z$ by using our relationship, $z=-t^2$. We see then that we have the powers $z^{-3/2}\bar{z}^{-3/2}$. If we go back to our original idea of the OPE and the fact that both the deformation and the SUGRA operators are (1,1), then this operator would be a power of $z^{h-1-1}=z^{-3/2}$ which implies $h=1/2$. So this operator is a (1/2,1/2) on the base. We note this is exactly the weight of the operator that mentioned as creating the leading singularity in equation (3.20) of \cite{Burrington:2012yq}.

Note that we did not explicitly include the contribution form the bare twist using the Lunin-Mathur technique. We want to relate the result of the OPE to something from the base. Since both sides of the OPE are lifted with the same map, they would have the exact same contribution when lifting. So we do not need to explicitly include it to find our results. This will no longer be true in the case of lifting the OPE of operators where both are in the twisted sector. We will include more details about this in the Discussion.

So we seem to have got the correct mixing operator, but it would be nice to check that this is the correct result. For this, we turn to the four point function, \\ $\langle\mathcal{O}_{dil}(a_1,\bar{a}_1) \mathcal{O}_D(b,\bar{b}) \mathcal{O}(0,0) \mathcal{O}_{dil}(a_2,\bar{a}_2)\rangle$, and consider the coincidence limit $(a_1,\bar a_1)\rightarrow(0,0)$ and $(a_2,\bar a_2)\rightarrow(b,\bar b)$. The leading singularity in the coincidence limit took the form of
\begin{equation}
\frac{2^{-4}}{|a_1|^3|a_2-b|^3|b|^{2}}
\end{equation}
This should be recoverable from our OPE. Our coefficient from the OPE is just 1. However, we must normalize the operators involved to get the structure constant correct. The normalization of the dilaton will give a factor of $(4)^{-1/2}$ and the deformation $(8)^{-1/2}$. Then we also need the normalization of the operator itself. Luckily, this is super easy to calculate. Consider,
\begin{equation}
\langle \epsilon_{\dot{A}\dot{B}}S^{\dot{A}\dot{B}}(1,1)\epsilon_{\dot{C}\dot{D}}S^{\dot{C}\dot{D}}(0,0)\rangle =\epsilon_{\dot{A}\dot{B}}\epsilon_{\dot{C}\dot{D}}\epsilon^{\dot{A}\dot{C}}\epsilon^{\dot{B}\dot{D}}=2
\end{equation}
So then we must multiply the operator by $2^{1/2}$ to get it properly normalized. Then we find
\begin{equation}
\frac{i}{|t|^6}\epsilon_{\dot{A}\dot{B}}S^{\dot{A}\dot{B}}=\frac{1}{z^{3/2}\bar{z}^{3/2}}(4)^{-1/2}(8)^{-1/2}2^{1/2}\mathcal{O}_{leading}=\frac{2^{-2}}{z^{3/2}\bar{z}^{3/2}}\mathcal{O}_{leading}
\end{equation}
where $\mathcal{O}_{leading}$ is the normalized mixing operator. So in the end we find we get the result $C_{dDl}=2^{-2}$.

To match up with the coefficient of the singularity, we must square this result. The leading singularity in the coincidence limit of the four point function would be the two point function of the normalized $\mathcal{O}_{leading}$ and together with its coefficient. So we would find that the coefficient of the leading singularity would be $C_{dDl}^2=2^{-4}$. This agrees with the coincidence limit of the four point function computed previously.

These results are relatively simple to work out. Now we turn to computing the mixing for a non-SUGRA mode where previous results had indicated there would be a non-trivial anomalous dimension.

\section{Results for candidate operator}\label{stringoperator}
\subsection{OPE and structure constants}

Let us consider our main candidate operator,
\begin{equation}
\mathcal{O}_C(z,\bar{z})=\epsilon_{AB}\epsilon_{\dot{A}\dot{B}}\epsilon_{EF}\epsilon_{\dot{E}\dot{F}}\sum\limits_k\partial X^{\dot{A}A}_{-1,(k)}\partial X^{\dot{B}B}_{-1,(k)}\bar{\partial}X^{\dot{E}E}_{-1,(k)}\bar{\partial}X^{\dot{F}F}_{-1,(k)},
\end{equation}
Operators that mix with it must be part of the twist 2 sector, as the three point function with the target operator and the deformation operator involves one twist 2 (from the deformation), the candidate mixing operator must have a complementary twist 2 to get an $S_N$ invariant correlator.

Our main approach is to analyze the OPE themselves, with the hope of finding a more efficient algorithm for computing mixing.  However, this approach will hit an immediate problem. We are looking for weight (2,2) quasi-primary operators for the computation of the anomalous dimension. In general, the result of the OPE will include a sum over quasi-primaries and their descendants. So we will need to find a way to project out the descendants. We will discuss a simple procedure to do this in Appendix \ref{projection}.

To show how this will help, let us consider doing the OPE of our candidate operator with the deformation operator. We will lift our OPE with the map $z=-t^2$, which will have the correct ramification of the twist 2 operator. Then we are looking at doing
\begin{equation}
\mathcal{O}_c(z,\bar{z}) \mathcal{O}_D(0,0)\rightarrow\sum\limits_{(t_{\pm},\bar{t}_{\pm})}\mathcal{O}_c(t_{\pm},\bar{t}_{\pm})\mathcal{O}^{(t)}_D(0,0)= 2\mathcal{O}_c(t,\bar{t})\mathcal{O}^{(t)}_D(0,0)
\end{equation}
Note here we have take the sum over images as just a simple multiple of 2. This is because the candidate operator just involves bosonic variables so the two images, which differ only by a sign, affects the odd orders of $t$ and will not matter in the end for our calculations. So to simplify the presentation, we will just skip that complication and give the OPE in terms of a generic $t$.

Moving forward, we note that lifting with our map the candidate operator lifts to
\begin{equation}
\mathcal{O}_c(t,\bar{t})=\frac{1}{16t^2\bar{t}^2}\left(\epsilon^{\dot{C}\dot{D}}\epsilon^{CD}\partial X_{\dot{C}C}(t)\partial X_{\dot{D}D}(t)+\frac{1}{t^2}\right)\left(\epsilon^{\dot{E}\dot{F}}\epsilon^{EF}\bar{\partial}X_{\dot{E}E}(\bar{t})\bar{\partial}X_{\dot{F}F}(\bar{t})+\frac{1}{\bar{t}^2}\right)
\end{equation}
and the deformation operator lifts to
\begin{equation}
\mathcal{O}^{(t)}_D(0,0)=-i\epsilon^{AB} \partial X_{\dot{A}A}\bar{\partial} X_{\dot{B}B}S^{\dot{A}\dot{B}}
\end{equation}
Note that in \cite{Burrington:2012yq} the authors had worked out that the deformation operator mixes with the candidate operator. Let's first show that we can reproduce this here. Since the deformation operator is weight $(1,1)$ and the candidate has $(2,2)$, we will be looking at order $z^{1-1-2}\bar{z}^{1-1-2}=z^{-2}\bar{z}^{-2}$ on the cover this corresponds to $t^{-4}\bar{t}^{-4}$) So then expanding out the OPE gives us
\begin{align}
&-\frac{i}{16t^2\bar{t}^2}\epsilon^{AB}\epsilon^{\dot{C}\dot{D}}\epsilon^{CD}\epsilon^{\dot{E}\dot{F}}\epsilon^{EF}\left(\frac{1}{t^2}\epsilon_{\dot{D}\dot{A}}\epsilon_{DA}\partial X_{\dot{C}C}(t)+\frac{1}{t^2}\epsilon_{\dot{C}\dot{A}}\epsilon_{CA}\partial X_{\dot{D}D}(t)+\frac{\epsilon_{\dot{C}\dot{D}}\epsilon_{CD}}{4t^2}\partial X_{\dot{A}A}\right)\\\nonumber
&\times\left(\frac{1}{\bar{t}^2}\epsilon_{\dot{F}\dot{B}}\epsilon_{FB}\bar{\partial}X_{\dot{E}E}(\bar{t})+\frac{1}{\bar{t}^2}\epsilon_{\dot{E}\dot{B}}\epsilon_{EB}\bar{\partial}X_{\dot{F}F}(\bar{t})+\frac{\epsilon_{\dot{E}\dot{F}}\epsilon_{EF}}{4\bar{t}^2}\bar{\partial} X_{\dot{B}B}\right)S^{\dot{A}\dot{B}}\\
&=-\frac{9i}{16t^4\bar{t}^4}\epsilon^{AB} \partial X_{\dot{A}A}\bar{\partial} X_{\dot{B}B}S^{\dot{A}\dot{B}}+...
\end{align}
Here we have suppressed the other terms in the OPE. We also want to include the normalization $2^{-4}$ for the candidate operator. So then we have in total
\begin{equation}
\mathcal{O}_c(z,\bar{z})\mathcal{O}_D(0,0)=2\times2^{-4}\times\frac{9}{2^4z^2\bar{z}^2}\mathcal{O}_D(0,0)+...=\frac{9}{2^7z^2\bar{z}^2}\mathcal{O}_D(0,0)+...
\end{equation}
This agrees with previous results, where the mixing of the deformation operator was found for the candidate as the leading singularity of the four point function. Recall the first factor of two is from the sum over images.

Next we can focus on looking for the mixing operator contributing to the anomalous dimension and see if the success holds. Again our OPE would yield the results
\begin{align}
&-\frac{i}{16t^2\bar{t}^2}\epsilon^{AB}\left(\frac{2}{t^2}\partial X_{\dot{A}A}(t)+\frac{1}{t^2}\partial X_{\dot{A}A}+\epsilon^{\dot{C}\dot{D}}\epsilon^{CD}\partial X_{\dot{C}C}(t)\partial X_{\dot{D}D}(t)\partial X_{\dot{A}A}\right)\\\nonumber
&\times\left(\frac{2}{\bar{t}^2}\bar{\partial} X_{\dot{B}B}(t)+\frac{1}{t^2}\bar{\partial}X_{\dot{B}B}+\epsilon^{\dot{E}\dot{F}}\epsilon^{EF}\bar{\partial}X_{\dot{E}E}(t)\bar{\partial}X_{\dot{F}F}(t)\bar{\partial}X_{\dot{B}B}\right)S^{\dot{A}\dot{B}}\\
&=\frac{i}{16t^2\bar{t}^2}\epsilon^{AB} (\partial X_{\dot{A}A}\epsilon^{\dot{C}\dot{D}}\epsilon^{CD}\partial X_{\dot{C}C}\partial X_{\dot{D}D}+\partial^3X_{\dot{A}A}) \nonumber\\
& \qquad \qquad \qquad \times (\bar{\partial} X_{\dot{B}B}\epsilon^{\dot{E}\dot{F}}\epsilon^{EF}\bar{\partial} X_{\dot{E}E}\bar{\partial}X_{\dot{F}F}+\bar{\partial}^3X_{\dot{B}B})S^{\dot{A}\dot{B}}+...
\end{align}
Again we have suppressed the other terms in the OPE. This is operator is not a quasi-primary. So we can use our procedure to find the quasi-primary contribution to the OPE at this order.

For the moment let us define some operator $\mathcal{A}$ which lifts on the cover to
\begin{align}
\mathcal{A}_{\uparrow}=&\epsilon^{AB}(\partial X_{\dot{A}A}\epsilon^{\dot{C}\dot{D}}\epsilon^{CD}\partial X_{\dot{C}C}\partial X_{\dot{D}D}+\partial^3X_{\dot{A}A})(\bar{\partial} X_{\dot{B}B}\epsilon^{\dot{E}\dot{F}}\epsilon^{EF}\bar{\partial}X_{\dot{E}E}\bar{\partial}X_{\dot{F}F}+\bar{\partial}^3X_{\dot{B}B})S^{\dot{A}\dot{B}}\nonumber\\
&=\epsilon^{AB}(\partial X_{\dot{A}A}\epsilon^{\dot{C}\dot{D}}\epsilon^{CD}\partial X_{\dot{C}C}\partial X_{\dot{D}D}+\partial^3X_{\dot{A}A})\tilde{Q}_{\dot{B}B}S^{\dot{A}\dot{B}}
\end{align}
We have suppressed the antiholomorphic parts for simplicity. We want to compute $(1-L_{-1}L_{1})\mathcal{A}$. Let us start with finding
\begin{align}
&L_1\mathcal{A}\rightarrow\\\nonumber
&\oint\frac{dt}{2\pi i}z(t)^2\left(\frac{dz}{dt}\right)^{-1}(T(t)-\frac{1}{2}\{z,t\})\epsilon^{AB}(\partial X_{\dot{A}A}\epsilon^{\dot{C}\dot{D}}\epsilon^{CD}\partial X_{\dot{C}C}\partial X_{\dot{D}D}+\partial^3X_{\dot{A}A})\tilde{Q}_{\dot{B}B}S^{\dot{A}\dot{B}}\\
=&-\frac{1}{2}\oint\frac{dt}{2\pi i}t^3T(t)\epsilon^{AB}(\partial X_{\dot{A}A}\epsilon^{\dot{C}\dot{D}}\epsilon^{CD}\partial X_{\dot{C}C}\partial X_{\dot{D}D}+\partial^3X_{\dot{A}A})\tilde{Q}_{\dot{B}B}S^{\dot{A}\dot{B}}
\end{align}
First note that the Schwarzian term will integrate to zero regardless, so we dropped this. Working out the OPE and taking the residue, we find
\begin{equation}
L_{1}\mathcal{A}\rightarrow -6\epsilon^{AB}\partial X_{\dot{A}A}\tilde{Q}_{\dot{B}B}S^{\dot{A}\dot{B}}
\end{equation}
Then we can look at
\begin{align}
&L_{-1}L_1\mathcal{A}\rightarrow\\\nonumber
&\oint\frac{dt}{2\pi i}\left(\frac{dz}{dt}\right)^{-1}(T(t)-\frac{1}{2}\{z,t\})( -6\epsilon^{AB}\partial X_{\dot{A}A}\tilde{Q}_{\dot{B}B}S^{\dot{A}\dot{B}})\\
=&\frac{3}{2}\oint\frac{dt}{2\pi i}t^{-1}T(t)\epsilon^{AB}\partial X_{\dot{A}A}\tilde{Q}_{\dot{B}B}S^{\dot{A}\dot{B}}
\end{align}
Again the Schwarzian does not contribute. We can work out the OPE, making sure that Taylor expand the singular terms of the OPE to correctly get the non singular terms. This will give us
\begin{align}
&L_{-1}L_1\mathcal{A}\rightarrow\\\nonumber
&\frac{3}{2}\epsilon^{AB}\Big(\partial X_{\dot{A}A} \epsilon^{\dot{C}\dot{D}}\epsilon^{CD}\partial X_{\dot{C}C}\partial X_{\dot{D} D} + \partial^3 X_{\dot{A}A}+2\partial X_{\dot{A} A} \partial^2 -\frac{1}{2}\partial X_{\dot{A}A}\left(\partial\phi^5-\partial\phi^6\right)^2\Big )\tilde{Q}_{\dot{B}B}S^{\dot{A}\dot{B}}
\end{align}
So then we find
\begin{align}
&(1-\frac{1}{2}L_{-1}L_1)\mathcal{A}\rightarrow\\\nonumber
&\frac{1}{8}\epsilon^{AB}\Big(2\partial X_{\dot{A}A} \epsilon^{\dot{C}\dot{D}}\epsilon^{CD}\partial X_{\dot{C}C}\partial X_{\dot{D} D} + 2\partial^3 X_{\dot{A}A}-12\partial X_{\dot{A} A} \partial^2 +3\partial X_{\dot{A}A}\left(\partial\phi^5-\partial\phi^6\right)^2\Big )\tilde{Q}_{\dot{B}B}S^{\dot{A}\dot{B}}
\end{align}

We now repeat the procedure with the antiholomorphic part of the operator as well. Putting this back into our OPE we find
\begin{align}
&\frac{i}{16t^2\bar{t}^2}\frac{1}{8}\frac{1}{8}\epsilon^{AB}\Big(2 \partial X_{\dot{A}A} \epsilon^{\dot{C}\dot{D}}\epsilon^{CD}\partial X_{\dot{C}C}\partial X_{\dot{D} D} +2 \partial^3 X_{\dot{A}A}-12\partial X_{\dot{A} A} \partial^2 +3\partial X_{\dot{A}A}\left(\partial\phi^5-\partial\phi^6\right)^2\Big ) \\\nonumber
& \Big(2\bar{\partial}X_{\dot{B}B} \epsilon^{\dot{E}\dot{F}}\epsilon^{EF}\bar{\partial}X_{\dot{E}E}\bar{\partial}X_{\dot{F}F} +2 \bar{\partial}^3 X_{\dot{B}B}-12\bar{\partial}X_{\dot{B}B} \bar{\partial}^2 +3\bar{\partial} X_{\dot{B}B}\left(\bar{\partial}\tilde{\phi}^5-\bar{\partial}\tilde{\phi}^6\right)^2\Big )S^{\dot{A}\dot{B}}\\
&=\frac{1}{2^{11}t^2\bar{t}^2}\times \nonumber \\
&\Bigg[2i\epsilon^{AB}\Big(2 \partial X_{\dot{A}A} \epsilon^{\dot{C}\dot{D}}\epsilon^{CD}\partial X_{\dot{C}C}\partial X_{\dot{D} D} +2 \partial^3 X_{\dot{A}A}-12\partial X_{\dot{A} A} \partial^2 +3\partial X_{\dot{A}A}\left(\partial\phi^5-\partial\phi^6\right)^2\Big ) \nonumber\\
& \Big(2\bar{\partial}X_{\dot{B}B} \epsilon^{\dot{E}\dot{F}}\epsilon^{EF}\bar{\partial}X_{\dot{E}E}\bar{\partial}X_{\dot{F}F} +2 \bar{\partial}^3 X_{\dot{B}B}-12\bar{\partial}X_{\dot{B}B} \bar{\partial}^2 +3\bar{\partial} X_{\dot{B}B}\left(\bar{\partial}\tilde{\phi}^5-\bar{\partial}\tilde{\phi}^6\right)^2\Big )S^{\dot{A}\dot{B}}\Bigg]
\end{align}
The operator in the brackets is the mixing operator we denote as $\mathcal{O}_{n2A}$. Now we can include the normalizations to find the structure constant. The candidate operator has the normalization $(2^8)^{-1/2}=2^{-4}$, the deformation $(2^3)^{-1/2}=2^{-3/2}$. The normalization of the mixing operator is $3\times2^{19/2}$, which was found by computing the two point function of $\mathcal{O}_{n2A}$ using our code. Together, the contributions work out to an extra factor of $3\times2^4$. We find that
\begin{equation}
\mathcal{O}_c(z,\bar{z})\mathcal{O}_D(0,0)=2\times(3\times2^4)\times\frac{1}{2^{11}z\bar{z}}\mathcal{O}_{n2A}(0,0)+...=\frac{3}{2^6z\bar{z}}\mathcal{O}_{n2A}(0,0)+...
\end{equation}
This result agrees with what we find by computing three point functions. In the next Section, we will discuss the details of the results of the three point functions worked out with our Mathematica package.

\subsection{Three point functions}

We undertook a search of quasi-primaries of weight (2,2) in the twist 2 sector for the three point function computations. The operators that will contribute to the anomalous dimension of our target operator are the quasi-primaries of weight (2,2) and singlets under the $SU(2)$s. This involves determining the effect of $L_0, \bar{L}_0$, to find the weight and then the effect of $L_{1},\bar{L}_{1}$ to see if the operator of interest would vanish as is expected for quasi-primaries. Of course, it is easier to check this on the cover, so we lift the computations to the cover where we would have no twists. Even with these constraints, there are still 34 operators that might participate in mixing. Doing this by hand would be very tedious, so we developed a  Mathematica code  to handle both the checking of operators being quasi-primary and the correlation functions. The length of calculations becomes considerably greater as we iterate the anomalous dimension procedure, and we will expand on this point in the Discussion.

From here we determine their three point functions to find our mixing operators. Not all of the operators that satisfy the restrictions will end up participate in mixing. In the end, we found that 9 operators contribute to mixing.
\begin{align}
\mathcal{O}_{1}&=\epsilon_{AE}\epsilon_{\dot{E}\dot{H}}\epsilon_{\dot{A}\dot{D}}\epsilon_{\alpha\beta}\epsilon_{\dot{\alpha}\dot{\beta}}\epsilon_{\dot{B}\dot{C}}\epsilon_{BC}\epsilon_{\dot{F}\dot{G}}\epsilon_{FG}\nonumber\\
&\partial X^{\dot{B}B}_{-1/2}\partial X^{\dot{C}C}_{-1/2}\partial X^{\dot{A}A}_{-1/2}\psi^{\alpha\dot{D}}_{0}\bar{\partial}X^{\dot{F}F}_{-1/2}\bar{\partial}X^{\dot{G}G}_{-1/2}\bar{\partial}X^{\dot{E}E}_{-1/2}\tilde{\psi}^{\dot{\alpha}\dot{H}}_{0}\sigma^{\beta\dot{\beta}}_2\\
\mathcal{O}_{2}&=\epsilon_{AE}\epsilon_{\dot{E}\dot{H}}\epsilon_{\dot{A}\dot{D}}\epsilon_{\alpha\beta}\epsilon_{\dot{\alpha}\dot{\beta}}\epsilon_{\dot{B}\dot{C}}\epsilon_{BC}\partial X^{\dot{B}B}_{-1/2}\partial X^{\dot{C}C}_{-1/2}\partial X^{\dot{A}A}_{-1/2}\psi^{\alpha\dot{D}}_{0}\bar{\partial}X^{\dot{E}E}_{-3/2}\tilde{\psi}^{\dot{\alpha}\dot{H}}_0\sigma^{\beta\dot{\beta}}_2\\
\mathcal{O}_{3}&=\epsilon_{AE}\epsilon_{\dot{E}\dot{H}}\epsilon_{\dot{A}\dot{D}}\epsilon_{\alpha\beta}\epsilon_{\dot{\alpha}\dot{\beta}}\epsilon_{\dot{B}\dot{C}}\epsilon_{BC}\partial X^{\dot{B}B}_{-1/2}\partial X^{\dot{C}C}_{-1/2}\partial X^{\dot{A}A}_{-1/2}\psi^{\alpha\dot{D}}_{0}\bar{\partial}X^{\dot{E}E}_{-1/2}\tilde{\psi}^{\dot{\alpha}\dot{H}}_{-1}\sigma^{\beta\dot{\beta}}_2\\
\mathcal{O}_{4}&=\epsilon_{AE}\epsilon_{\dot{E}\dot{H}}\epsilon_{\dot{A}\dot{D}}\epsilon_{\alpha\beta}\epsilon_{\dot{\alpha}\dot{\beta}}\epsilon_{\dot{F}\dot{G}}\epsilon_{FG}\partial X^{\dot{A}A}_{-3/2}\psi^{\alpha\dot{D}}_0\bar{\partial}X^{\dot{F}F}_{-1/2}\bar{\partial}X^{\dot{G}G}_{-1/2}\bar{\partial}X^{\dot{E}E}_{-1/2}\tilde{\psi}^{\dot{\alpha}\dot{H}}_{0}\sigma^{\beta\dot{\beta}}_2\\
\mathcal{O}_{5}&=\epsilon_{AE}\epsilon_{\dot{E}\dot{H}}\epsilon_{\dot{A}\dot{D}}\epsilon_{\alpha\beta}\epsilon_{\dot{\alpha}\dot{\beta}}\partial X^{\dot{A}A}_{-3/2}\psi^{\alpha\dot{D}}_0\bar{\partial}X^{\dot{E}E}_{-3/2}\tilde{\psi}^{\dot{\alpha}\dot{H}}_{0}\sigma^{\beta\dot{\beta}}_2\\
\mathcal{O}_{6}&=\epsilon_{AE}\epsilon_{\dot{E}\dot{H}}\epsilon_{\dot{A}\dot{D}}\epsilon_{\alpha\beta}\epsilon_{\dot{\alpha}\dot{\beta}}\partial X^{\dot{A}A}_{-3/2}\psi^{\alpha\dot{D}}_0\bar{\partial}X^{\dot{E}E}_{-1/2}\tilde{\psi}^{\dot{\alpha}\dot{H}}_{-1}\sigma^{\beta\dot{\beta}}_2\\
\mathcal{O}_{7}&=\epsilon_{AE}\epsilon_{\dot{E}\dot{H}}\epsilon_{\dot{A}\dot{D}}\epsilon_{\alpha\beta}\epsilon_{\dot{\alpha}\dot{\beta}}\epsilon_{\dot{F}\dot{G}}\epsilon_{FG}\partial X^{\dot{A}A}_{-1/2}\psi^{\alpha\dot{D}}_{-1}\bar{\partial}X^{\dot{F}F}_{-1/2}\bar{\partial}X^{\dot{G}G}_{-1/2}\bar{\partial}X^{\dot{E}E}_{-1/2}\tilde{\psi}^{\dot{\alpha}\dot{H}}_{0}\sigma^{\beta\dot{\beta}}_2\\
\mathcal{O}_{8}&=\epsilon_{AE}\epsilon_{\dot{E}\dot{H}}\epsilon_{\dot{A}\dot{D}}\epsilon_{\alpha\beta}\epsilon_{\dot{\alpha}\dot{\beta}}\partial X^{\dot{A}A}_{-1/2}\psi^{\alpha\dot{D}}_{-1}\bar{\partial}X^{\dot{E}E}_{-3/2}\tilde{\psi}^{\dot{\alpha}\dot{H}}_{0}\sigma^{\beta\dot{\beta}}_2\\
\mathcal{O}_{9}&=\epsilon_{AE}\epsilon_{\dot{E}\dot{H}}\epsilon_{\dot{A}\dot{D}}\epsilon_{\alpha\beta}\epsilon_{\dot{\alpha}\dot{\beta}}\partial X^{\dot{A}A}_{-1/2}\psi^{\alpha\dot{D}}_{-1}\bar{\partial}X^{\dot{E}E}_{-1/2}\tilde{\psi}^{\dot{\alpha}\dot{H}}_{-1}\sigma^{\beta\dot{\beta}}_2
\end{align}
 However, these are not quasi-primaries on their own. So we take a linear combination of them, $\sum_ia_i\mathcal{O}_i$. Applying the $L_1$ operator to the linear combination and setting the equation to zero lead the equations $6a_1+3a_4+a_7=0$, $6a_2+3a_5+a_8=0$, and $6a_3+3a_6+a_9=0$. Applying the $\bar{L}_1$ operator to the linear combination and setting the equation to zero lead the equations $6a_1+3a_2+a_3=0$, $6a_4+3a_5+a_6=0$, and $6a_7+3a_8+a_9=0$. These equations will lead to four independent solutions. We originally considered the combination,
 \begin{align}
\mathcal{O}_{11}&=\epsilon_{AE}\epsilon_{\dot{E}\dot{H}}\epsilon_{\dot{A}\dot{D}}\epsilon_{\alpha\beta}\epsilon_{\dot{\alpha}\dot{\beta}}(\epsilon_{\dot{B}\dot{C}}\epsilon_{BC}\partial X^{\dot{B}B}_{-1/2}\partial X^{\dot{C}C}_{-1/2}\partial X^{\dot{A}A}_{-1/2}\psi^{\alpha\dot{D}}_{0}-6\partial X^{\dot{A}A}_{-1/2}\psi^{\alpha\dot{D}}_{-1})\nonumber\\
&(\epsilon_{\dot{F}\dot{G}}\epsilon_{FG}\bar{\partial}X^{\dot{F}F}_{-1/2}\bar{\partial}X^{\dot{G}G}_{-1/2}\bar{\partial}X^{\dot{E}E}_{-1/2}\tilde{\psi}^{\dot{\alpha}\dot{H}}_{0}- 6\bar{\partial}X^{\dot{E}E}_{-1/2}\tilde{\psi}^{\dot{\alpha}\dot{H}}_{-1})\sigma^{\beta\dot{\beta}}_2\\
\mathcal{O}_{12}&=\epsilon_{AE}\epsilon_{\dot{E}\dot{H}}\epsilon_{\dot{A}\dot{D}}\epsilon_{\alpha\beta}\epsilon_{\dot{\alpha}\dot{\beta}}(\epsilon_{\dot{B}\dot{C}}\epsilon_{BC}\partial X^{\dot{B}B}_{-1/2}\partial X^{\dot{C}C}_{-1/2}\partial X^{\dot{A}A}_{-1/2}\psi^{\alpha\dot{D}}_{0}-6\partial X^{\dot{A}A}_{-1/2}\psi^{\alpha\dot{D}}_{-1})\nonumber\\
&(\bar{\partial}X^{\dot{E}E}_{-3/2}\tilde{\psi}^{\dot{\alpha}\dot{H}}_0-3\bar{\partial}X^{\dot{E}E}_{-1/2}\tilde{\psi}^{\dot{\alpha}\dot{H}}_{-1})\sigma^{\beta\dot{\beta}}_2\\
\mathcal{O}_{21}&=\epsilon_{AE}\epsilon_{\dot{E}\dot{H}}\epsilon_{\dot{A}\dot{D}}\epsilon_{\alpha\beta}\epsilon_{\dot{\alpha}\dot{\beta}}(\partial X^{\dot{A}A}_{-3/2}\psi^{\alpha\dot{D}}_0-3\partial X^{\dot{A}A}_{-1/2}\psi^{\alpha\dot{D}}_{-1})\nonumber\\
&(\epsilon_{\dot{F}\dot{G}}\epsilon_{FG}\bar{\partial}X^{\dot{F}F}_{-1/2}\bar{\partial}X^{\dot{G}G}_{-1/2}\bar{\partial}X^{\dot{E}E}_{-1/2}\tilde{\psi}^{\dot{\alpha}\dot{H}}_{0}- 6\bar{\partial}X^{\dot{E}E}_{-1/2}\tilde{\psi}^{\dot{\alpha}\dot{H}}_{-1})\sigma^{\beta\dot{\beta}}_2\\
\mathcal{O}_{22}&=\epsilon_{AE}\epsilon_{\dot{E}\dot{H}}\epsilon_{\dot{A}\dot{D}}\epsilon_{\alpha\beta}\epsilon_{\dot{\alpha}\dot{\beta}}(\partial X^{\dot{A}A}_{-3/2}\psi^{\alpha\dot{D}}_0-3\partial X^{\dot{A}A}_{-1/2}\psi^{\alpha\dot{D}}_{-1})(\bar{\partial}X^{\dot{E}E}_{-3/2}\tilde{\psi}^{\dot{\alpha}\dot{H}}_0-3\bar{\partial}X^{\dot{E}E}_{-1/2}\tilde{\psi}^{\dot{\alpha}\dot{H}}_{-1})\sigma^{\beta\dot{\beta}}_2
\end{align}

Now we can consider the operator $\mathcal{O}=c_{11}\mathcal{O}_{11}+c_{12}\mathcal{O}_{12}+c_{21}\mathcal{O}_{21}+c_{22}\mathcal{O}_{22}$. We wish to use this operator in the three point function. To compute the three point function, $\langle \mathcal{O}_C(a_1,\bar{a}_1)\mathcal{O}_d(b,\bar{b})\mathcal{O}(0,0)\rangle$, we will lift to the covering space with the map
\begin{equation}\label{mapfor3pf}
z(t)=\frac{bt^2}{2t-1}
\end{equation}
This map has the correct ramifications for twist 2 operators around $z=b,t=1$ and $z=0,t=0$. Lifting to the cover and computing the three point function with the Lunin-Mathur technique, we find
\begin{equation}
\langle \mathcal{O}_C(a_1,\bar{a}_1)\mathcal{O}_D(b,\bar{b})\mathcal{O}(0,0)\rangle=\frac{288(4c_{11}+2c_{12}+2c_{21}+c_{22})}{a_1^3(a_1-b)b\bar{a}_1^3(\bar{a}_1-\bar{b})\bar{b}}.
\end{equation}
This result includes the contributions from the bare twists. However, our original set of operators turn out to not be orthogonal. The orthogonal combinations can be obtained by the usual Gram-Schmidt procedure. Doing so will give the results,
\begin{align}
\tilde{\mathcal{O}}_{11}&=\mathcal{O}_{11}\\
\tilde{\mathcal{O}}_{12}&=\mathcal{O}_{12}-\frac{3}{10}\mathcal{O}_{11}\\
\tilde{\mathcal{O}}_{21}&=\mathcal{O}_{21}-\frac{3}{10}\mathcal{O}_{11}\\
\tilde{\mathcal{O}}_{22}&=\mathcal{O}_{22}-\frac{3}{10}\mathcal{O}_{21}-\frac{3}{10}\mathcal{O}_{12}+\frac{9}{100}\mathcal{O}_{11}.
\end{align}
We still need to normalize this by the square root of the two point function of the mixing operator, $\langle\mathcal{O}(1,1)\mathcal{O}(0,0)\rangle$. Computing this can be done with the same map in eq.(\ref{mapfor3pf}), where we simply set $b=1$. We then find
\begin{align}\label{norm}
\langle\mathcal{O}(1,1)\mathcal{O}(0,0)\rangle&=460800 c_{11}^2 + 276480 c_{11}c_{21} + 115200 c_{21}^2 + 276480 c_{11} c_{12}\\\nonumber
& + 165888 c_{11}c_{22} + 69120 c_{21}c_{22} + 115200 c_{12}^2 + 69120 c_{12}c_{22} +28800 c_{22}^2
\end{align}

Using eq.(\ref{norm}) to normalize our mixing operator together with the normalizations of the two point functions of the deformation, $\langle\mathcal{O}_D(1,1)\mathcal{O}_D(0,0)\rangle=8$ and the candidate, \\ $\langle\mathcal{O}_C(1,1)\mathcal{O}_C(0,0)\rangle=2^8$, will lead to the four results,
\begin{align}
\langle \mathcal{O}_C(a_1,\bar{a}_1)\mathcal{O}_D(b,\bar{b})\tilde{\mathcal{O}}_{11}(0,0)\rangle=&\frac{3}{80a_1^3\bar{a}_1^3(a_1-b)(\bar{a}_1-\bar{b})b\bar{b}}\\
\langle \mathcal{O}_C(a_1,\bar{a}_1)\mathcal{O}_D(b,\bar{b})\tilde{\mathcal{O}}_{12}(0,0)\rangle=& \frac{3}{160a_1^3\bar{a}_1^3(a_1-b)(\bar{a}_1-\bar{b})b\bar{b}}\\
\langle \mathcal{O}_C(a_1,\bar{a}_1)\mathcal{O}_D(b,\bar{b})\tilde{\mathcal{O}}_{21}(0,0)\rangle=&\frac{3}{160a_1^3\bar{a}_1^3(a_1-b)(\bar{a}_1-\bar{b})b\bar{b}}\\
\langle \mathcal{O}_C(a_1,\bar{a}_1)\mathcal{O}_D(b,\bar{b})\tilde{\mathcal{O}}_{22}(0,0)\rangle=&\frac{3}{320a_1^3\bar{a}_1^3(a_1-b)(\bar{a}_1-\bar{b})b\bar{b}}.
\end{align}
We can simplify our expression by changing our basis of operators here to have only one operator that accounts for the full mixing and we can also normalize it. This operator that captures the mixing is given by
\begin{equation}
\mathcal{O}_{n2A}=(\mathcal{O}_{11}+2\mathcal{O}_{12}+2\mathcal{O}_{21}+4\mathcal{O}_{22})/(1536\sqrt{2}).
\end{equation}
The structure constant for this operator is
\begin{equation}
C_{iD(n2A)}=\frac{3}{64}.
\end{equation}
Both the operator and structure constant agree with the result obtained from the OPE.

\subsection{Four point function: coincidence limit}

A non-trivial check on both methods is to examine the four point function of two candidates and two deformations. The coincidence limit will indicate the mixing for the operator. The four point function for the candidate operator had previously been worked out,
\begin{align}
&\langle\mathcal{O}_C(a_1,\bar a_1)\,\mathcal{O}_{D}(b,\bar b)\mathcal{O}_{D}(0,\bar 0)\mathcal{O}_C(a_2,\bar a_2)\rangle=\frac{|b|^4}{2^{12}}\lambda^2|a_1|^{-4}|a_2|^{-4}|a_1-b|^{-4}|a_2-b|^{-4}\bigg\{\frac{25}{4}\nonumber\\
&+5\left(\frac{(R+1)^2}{(R-1)^2}+\frac{(\bar{R}+1)^2}{(\bar{R}-1)^2}\right)+16\frac{R^{1/2}(R+1)}{(R-1)^2}\frac{\bar{R}^{1/2}(\bar{R}+1)}{(\bar{R}-1)^2}+4(\frac{(R+1)^2}{(R-1)^2}\frac{(\bar{R}+1)^2}{(\bar{R}-1)^2}\nonumber\\
&+10\left(\frac{R(R^2+6R+1)}{(R-1)^4}+\frac{\bar{R}(\bar{R}^2+6\bar{R}+1)}{(\bar{R}-1)^4}\right)+16\frac{R(R^2+6R+1)}{(R-1)^4}\frac{\bar{R}(\bar{R}^2+6\bar{R}+1)}{(\bar{R}-1)^4}\nonumber\\
&+64\left(\frac{R^{1/2}(R+1)}{(R-1)^2}\frac{\bar{R}^{3/2} (\bar{R}+1)}{(\bar{R}-1)^4}+\frac{R^{3/2}(R+1)}{(R-1)^4}\frac{\bar{R}^{1/2}(\bar{R}+1)}{(\bar{R}-1)^2}\right) \nonumber \\
& +256\frac{R^{3/2}(R+1)}{(R-1)^4}\frac{\bar{R}^{3/2}(\bar{R}+1)}{(\bar{R}-1)^4} \nonumber \\
&+8\left(\frac{(R+1)^2}{(R-1)^2}\frac{\bar{R}(\bar{R}^2+6\bar{R}+1)}{(\bar{R}-1)^4} +\frac{R(R^2+6R+1)}{(R-1)^4}\frac{(\bar{R}+1)^2}{(\bar{R}-1)^2}\right)\bigg\},
\end{align}
where $R=a_1(a_2-b)(a_2(a_1-b))$.

We wish to look at the coincidence limit, $(a_1,\bar a_1)\rightarrow(0,0)$ and $(a_2,\bar a_2)\rightarrow(b,\bar b)$. There will be an number of singularities in this limit, but we are only interested in the mixing from a weight (2,2), as mixing with other operators will not contribute to the anomalous dimension. To see what comes from mixing, one first has to look at (1,1) operators that contribute and subtract the weight (2,2) descendants' contribution, as mixing will only come from quasi-primaries not descendants. In the previous work \cite{Burrington:2012yq}, it was found that the only operator at weight (1,1) was the deformation operator itself. We also found this using our OPE method. Subtracting its descendant's contribution, the term that indicates mixing with quasi-primary operators is given by
\begin{equation}\label{MixSingu}
\frac{9}{2^{12}a_1(a_2-b)\bar{a_1}(\bar{a}_2-\bar{b})|b|^8}
\end{equation}
This is the singularity that indicates the mixing with weight (2,2) operators. Using  \\ $C_{iD(n2A)}=3/2^{6}$, we find that $C_{iD(n2A)}^2=9/2^{12}$ gives us the coefficient in the four point function. This indicates we have found the correct and complete mixing at twist 2.

\section{Discussion}\label{discussion}

To fully compute the anomalous dimension, we would need to consider computing the mixing for higher twist operators. This is due to the fact that a twist 2 operator can mix with a twist 3 under the deformation and so on. Fixing the conformal dimension will put a cutoff on how high the twists can go, as the bare twist fields carry conformal weight themselves. A simple calculation will show that $n=8$ is the highest twist allowed for a $(2,2)$ operator.

Doing this procedure for higher twists using correlation functions is well defined. However, due to the fractional moding of operators, the number of operators at each twist level will generally increase. This means computing increasing numbers of correlators, which will also have more complicated structure. In total, we believe there would be a total of 288, including the candidate operator, that might be involved in the mixing. The code we have developed can do these computations as well, but it takes an unreasonably long time. Our preliminary calculations with the code showed that there were 9 independent mixing operators at twist 3 and 36 independent mixing operators at twist 4, and twist 5 was just as complicated. Finishing the computations by brute force was unappetizing owing to the enormous length of expressions obtained.

To avoid this problem, we would like to use our OPE algorithm to find the higher twist operators that participate in mixing. Then we would would immediately recover the mixing operator from the one computation. However, this would require understanding how to properly lift the OPE to the cover when there are two twisted operators, and it turns out that there is a really interesting subtlety to this story. On the one side, we would have an OPE like
\begin{equation}
\mathcal{O}_D(z_0,\bar{z}_0)\mathcal{O}(0,0)
\end{equation}
where $\mathcal{O}_D$ is the deformation operator and $\mathcal{O}$ is the operator we are investigating the mixing for (which for our candidate is not in the twisted sector but in general may be a twisted operator with twist $n$). To get the ramifications correct when lifting this, we could use the map
\begin{equation}
z=-nt^n\left(t-\frac{n+1}{n}t_0\right)
\end{equation}
This has the correct ramification for the twist 2 deformation operator at $t=t_0$ and the correct ramification for a twist $n$ operator at $t=0$. Note that $z_0=t_0^{n+1}$, so the image of $z_0$ becomes $e^{2\pi i k/(n+1)}t_0$ on the cover. A sum over these will ensure that will have correct power series expansion of $z_0$. However, the end result of the OPE would be an operator of twist $n+1$ at $t=0$, which the map does not correctly ramify. We could consider the limit of the map $t_0\rightarrow0$, which would give us a map $z=-nt^{n+1}$. This has the correct ramification.  However, in this process, a ramified point in the map has been moved.  In the Lunin-Mathur covering space technique, there are extra contributions coming from a Liouville action contribution \cite{Lunin:2000yv}, and these contributions are localized at ramified points in the covering space. We checked this problem by computing the $\mathcal{O}_D(z_0,\bar{z}_0)\mathcal{O}_C(0,0)$ OPE and found that, not being careful about the map changing leads to an incorrect result.

Another way to see the same difficulty is looking at the action of translation on a generic operator.  At leading order,
\begin{equation}
\mathcal{O}(z_0+\delta z_0)=\mathcal{O}(z_0)+\partial\mathcal{O}(z_0)\delta z_0+\cdots=\mathcal{O}(z_0)+(L_{-1}\mathcal{O})(z_0)\delta z_0+\cdots
\end{equation}
The change $(L_{-1}\mathcal{O})(z_0)\delta z_0$ may be lifted to the covering surface by a map $z=f(t)$ where $z_0=f(t_0)$ for some $t_0$, and we find
\begin{equation}
(L_{-1}\mathcal{O})(z_0)\rightarrow \oint_{t_0} \frac{dt}{2\pi i} \left(\frac{\partial f(t)}{\partial t}\right)^{-1} \left(T(t) -\frac{c}{12}\left\{f(t),t\right\}\right) \mathcal O_{\uparrow} (t_0)
\end{equation}
We see here that if the map is ramified at $t=t_0$ with ramification $n$ the function $\partial f(t)/\partial t$ will have an $n^{\rm th}$ order 0 at $t_0$, leading to Virasoro generators $L_{-k}$ with $k\geq 2$ on the cover.  Thus, simple translation on the base space does not relate so easily to simple translation on the cover.  Simple translation on the cover is only recovered for non-ramified points in the map, i.e. only applies to untwisted operators.

We plan to return to this issue in a future paper. We still expect that the knowledge of the mixing is contained in the OPE on the cover, along with some additional information coming from a careful treatment of the Liouville dressing terms.

Another point to keep in mind is that the correlators and structure constants computed here were done using a representative correlator. To get a fully $S_N$ invariant correlator in the orbifold CFT, we would need to sum over the conjugacy classes of the representative twist $\sigma_{(12)}$. This would lead to some combinatorial factors that we have suppressed thusfar using a particular prescription for stripping these terms.  To avoid the complications of these combinatorial factors, and to compare to the results of earlier work, we chose to follow the conventions of \cite{Burrington:2012yq}.

There are a number of interesting further potential applications of this D1D5 technology. The most obvious would be to look at other operators than the candidate studied here. However, this would involve larger numbers of possible operators in the mixing as one increases the conformal weight. This would make the three point function searches unfeasible but one could study the OPE of any untwisted and twisted operator just as easily as we have here. Going further, one could extend the methods here used for compute mixing to computing the general structure constants of other operators with other perturbations of the D1D5 system.

Another application would be to continue to study the deformation from the point of view of higher spin theory. Previous work has shown that at the D1D5 CFT at the orbifold point has a subsector of Vasiliev higher spin theory \cite{Gaberdiel:2014cha}\cite{Gaberdiel:2015mra}\cite{Gaberdiel:2015wpo}. Then the deformation away from the orbifold point acts like as a Higgs mechanism, giving mass to these higher spin fields which are not present in the supergravity description. There has been work shown on finding the anomalous dimensions of the generators of these higher spin symmetries \cite{Gaberdiel:2015uca}. It would be interesting to see how using OPEs might help explore this further, or how the OPE method fits into the context of $W_{\infty}$ representation theory.

There are other recent works done using D1D5 technology that might be able to be taken further with our code and/or our methods. One can consider a toy model \cite{Lunin:2012gz} of infall in this D1D5 system. It may be interesting to study other such infalls using more complicated states (such as ones in \cite{Mathur:2011gz} or using the duals of arbitrary angular momenta 3-charge system \cite{Bena:2016ypk}) other than the Ramond vacuum. This would involve computing complicated correlators, which our code should be able to accomplish.

There has also been a large amount of work done studying the conformal blocks in CFT \cite{Hijano:2015zsa}\cite{Hijano:2015qja} and OPE blocks using kinematic space \cite{Czech:2015qta}\cite{Czech:2016xec}. Since we have a free theory, we can calculate these conformal blocks directly. It might be interesting to show some explicit examples of these new methods for calculating conformal blocks, especially if they could be used to generate conformal blocks and by extension correlators in the D1D5 CFT more efficiently than the methods presented here.

As we continue to explore the D1D5 CFT, we are hoping that new tools will allow us to come to a better understanding of this important model.

\section*{Acknowledgements}
We thank Ida G. Zadeh for collaboration in the early stages of this project.
The work of AWP and ITJ is supported by a Discovery Grant from the Natural Sciences and Engineering Research Council of Canada. The work of BAB is supported by funds provided by Hofstra University, including
a Faculty Research and Development Grant, and faculty startup funds.  Computations were performed on the GPC supercomputer at the SciNet HPC Consortium. SciNet is funded by: the Canada Foundation for Innovation under the auspices of Compute Canada; the Government of Ontario; Ontario Research Fund - Research Excellence; and the University of Toronto.

\appendix
\section{Quasi-primary projection procedure}\label{projection}

Start with a state $\mid\phi_n\rangle$ with conformal weight $h$ and suppose the state satisfies
\begin{equation}
L_1^{n+1}\mid\phi_n\rangle=0,L_1^{n}\mid\phi_n\rangle\neq0
\end{equation}
Now construct the following state
\begin{equation}
\mid \phi_{n-1}\rangle=\left(1-\frac{1}{N(n)} L_{-1}L_1\right)\mid\phi_n\rangle
\end{equation}
with
\begin{equation}
N(n)=n(2h-(n+1))
\end{equation}
One can show that the state $\mid \phi_{n-1}\rangle$ satisfies $L_1^n\mid \phi_{n-1}\rangle=0$.
We may iterate this process to obtain
\begin{equation}
\mid\psi\rangle=\left(1-\frac{1}{N(1)}L_{-1}L_1\right)\left(1-\frac{1}{N(2)}L_{-1}L_1\right)\cdots\left(1-\frac{1}{N(n)} L_{-1}L_1\right)\mid\phi_n\rangle
\end{equation}
where $L_1\mid \psi \rangle=0$, and so $\mid \psi \rangle$ is quasi primary. So in general, this will be the approach we will take to get a quasi-primary out of a non-QP.


\end{document}